# Cellulose-Based Reflective Liquid Crystal Films as Optical Filters and Solar Gain Regulators

Joshua A. De La Cruz,[†] Qingkun Liu,[‡] Bohdan Senyuk,[‡] Allister W. Frazier,[†] Karthik Peddireddy,[‡] and Ivan I. Smalyukh[*,†,‡,§,⊥]

[†]Materials Science and Engineering Program, University of Colorado, Boulder, Colorado 80309, United States

[‡]Department of Physics, University of Colorado, Boulder, Colorado 80309, United States

[§]Soft Materials Research Center and Department of Electrical, Computer and Energy Engineering, University of Colorado, Boulder, Colorado 80309, United States

[⊥]Renewable and Sustainable Energy Institute, National Renewable Energy Laboratory and University of Colorado, Boulder, Colorado 80303, United States

**ABSTRACT**: Many promising approaches for designing interactions of synthetic materials with light involve solid optical monocrystals and nanofabricated photonic crystal structures with spatially periodic variations of refractive index. Although their high costs limit current technological applications, remarkably, such photonic and optically anisotropic materials have also evolved throughout nature and enable narrow or broad-band spectral reflection of light. Here we use self-assembly of biomaterial cellulose nanocrystals to obtain three-layer films with helicoidal and nematic-like organization of the cellulose nanoparticles, which mimics naturally occurring polarization-insensitive reflectors found in the wings of *Plusiotis resplendens* beetles. These films were characterized with polarized optical microscopy and circular dichroism spectrometry, as well as scanning and transmission electron microscopies. These films exhibit high reflectivity tunable within the visible and near-infrared regions of the optical spectrum and may find applications ranging from color filters to smart cloth designs and in solar-gain-regulating building technologies.

**KEYWORDS**: photonic crystal, circular polarizer, tunable reflection, chiral nematic, biomaterials, liquid crystal

Photonic crystals enable a host of practical applications such as nonlinear waveguides,[1,2] advanced anticounterfeiting structures,[3,4] and low-energy-consuming displays;[5,6] however, the bulk of these material systems are carefully prepared via top-down methods such as photo- and electron-lithography. Nanofabrication affords precise control over structure but is obtained at the cost of detrimentally impacting scalability and throughput, hampering economic feasibility, and ultimately limiting such materials' practical scope. Nonetheless, photonic crystals produced bottom up via self-assembly are abundant throughout biology, including those in aberration-corrected spherical lenses found in the arm ossicles of the marine organism *Ophiocoma wendtii* and viewing-angle-independent selective reflecting exoskeletons of the beetle species *Plusiotis resplendens*.[7–9] These self-assembled photonic structures have driven researchers to apply bottom-up manufacturing in developing other synthetic photonic crystal systems, resulting in the formation of novel synthetic photonic material systems such as photonic diodes,[9] tunable organic lasers,[10] and subpicosecond optical switches.[11] The cost of these bottom-up crystal systems can further be decreased by utilizing common biomaterials such as cellulose. Cellulose is the most abundant biopolymer on the planet and is composed of polymerized D-glucose dimer chains arranged in a semicrystalline fashion.[12] Cellulose nanocrystals (CNCs) are particularly interesting for photonic applications, as cellulose is an inexpensive renewable resource that can self-assemble into periodic optical planes serving as Bragg diffractors.[13] However, the theoretical optical performance of traditional CNC-based systems is limited to 50% due to selective reflection of like-handed polarized radiation over opposite-handed polarized radiation,[14] limiting their optical applications. This limitation thus far has only been overcome with the use of non-biological-based retardation materials such as pentylcyanobiphenyl (5CB).[9]

In this work, we present the preparation, characterization, and optical modeling of a transparent, flexible, and economical CNC-based photonic composite structure serving as polarization-independent tunable Bragg reflectors operating within the visible and near-infrared (near-IR) regimes. CNCs derived from both bacteria cellulose and cotton cellulose are utilized in this study. The polarization-independent reflection of the laminar composite structure detailed herein is realized by insertion of a birefringent half-wave plate composed of nematic-like CNC between two reflective cholesteric-like CNC−organosilica composite films, forming a laminar composite "sandwich structure" mimicking the exoskeletal of the aforementioned beetles.[15] Electron microscopies reveal the spindle-like geometry of CNCs, and polarized optical microscopy shows their adoption of both nematic-like and cholesteric-like assemblies within the solid films, as controlled by our sample preparation conditions. The optical performance of the sandwich structure is contrasted with that of a single cholesteric-like film via reflective circular dichroism (CD) spectrometry and polarized optical microscopy (POM). The structure of the three layers is determined via scanning and transmission electron microscopies. Reflection spectra reveal a ∼95% peak reflectivity increase compared to a single film, with the total peak reflectivity of the structure reaching as high as 82%. These CNC-based laminar composite photonic structures may enable applications such as low-emissivity and solar-gain-regulating films, color and transparent visible-infrared filters, smart fabrics, optically enabled document security features, optical isolators, and other applications where shielding radiative heat while preserving visible-range transparency is needed. Furthermore, the use of cheap, abundant biomaterials such as cellulose and the implementation of a bottom-up fabrication approach make these materials a feasible, cost-effective solution that is readily implemented.

**RESULTS**

**Design and Physical Principles.** A single left-handed cholesteric-like CNC film's theoretical 50% maximum reflection limit due to selective reflection is overcome by construction of a highly reflective and flexible composite CNC−organosilica photonic structure, whose schematic is detailed in Figure 1a. The proposed photonic structure consists of two left-handed cholesteric-like composite cotton-derived CNC−organosilica films sandwiching an aligned nematic-like bacterial CNC optical retarder and possesses a polarization-independent reflection of incident radiation satisfying the Bragg diffraction condition.[16] The three-layered approach to achieve total reflection from solely left-handed films is inspired by the beetle species *Plusiotis resplendens*, whose exoskeleton is composed of two left-handed helicoidal layers sandwiching a nematic-like assembly; all three of these layers are primarily composed of the same polysaccharide, chitin, the second most abundant biopolymer after cellulose.[17] CNCs serve as a robust templating platform, imparting the cholesteric assembly onto soluble guest chemical species introduced to the colloidal dispersion,13 allowing the film's optical, mechanical, and chemical properties to be tailored; the cholesteric-like films' chemistry is optimized to maximize reflection, and the plasticizer polyethylene glycol ($M_w$: 400 g/mol, PEG-400) is added to improve flexibility and mechanical toughness. Unidirectional shearing can force CNCs' adoption of a nematic-like assembly, forming macroscopically aligned monodomain films, which serve as a birefringent phase retarder with designed retardation via a series of successive linear shear deposition cycles.

The polarization-independent reflection of the sandwich structure results from the combination of cholesteric- and nematic-like CNC optical elements, enabling reflection of both incident left-handed circularly polarized (LCP) light and righthanded circularly polarized (RCP) light solely by left-handed cholesteric-like CNC films. When natural light is incident upon the sandwich structure, the LCP component of light corresponding to the Bragg diffraction condition is selectively reflected. As incident RCP light is transmitted through the first cholesteric-like CNC film, it is converted into LCP light by the nematic-like CNC retarder acting as a half-wave plate. This LCP light is reflected by the second cholesteric-like layer and is converted back into RCP light as it transmits through the retarder again. Finally, this RCP radiation transmits through the first left-handed cholesteric-like film. Thus, we see all polarizations of light are reflected, as depicted in the scheme in Figure 1b. The reflective performance of the structure is polarization-independent and shows 2-fold reflection enhancement as compared to a single cholesteric-like CNC film. Because the pitch of the cholesteric-like films and the retardation of the nematic-like film are tunable from the visible to near-IR regimes, it follows that this reflection enhancement is also tunable in those same regimes.

The flexible cholesteric-like CNC films' arbitrarily tunable reflection within the visible and near-IR regimes is achieved by controlling synthetic and processing conditions. Sulfuric-acidcatalyzed hydrolysis of cellulose preferentially etches amorphous cellulose domains, yielding colloidal chiral cellulose nanocrystalline spindle-shaped rods.[18] Different cellulose sources vary in their degree of crystallinity and size of crystallites,[16] and thus the size and yield of CNCs obtained strongly depend on their source. Colloidal stability is ensured by Coulombic repulsion between the negatively charged sulfate ester moieties of the CNCs. Concentrating above the CNC's critical Onsager concentration (about 3.5 wt % for bacterial CNCs and 5.0 wt % for cotton CNCs, depending on synthetic conditions and processing) results in a thermodynamically stable lyotropic liquid crystal (LC).[19] The CNCs' inherit geometric chirality leads to adoption of the left-handed chiral nematic (cholesteric) LC phase with a helical axis denoted by $\chi$ (Figure 1b).[20,21] The cholesteric phase's twisting repeated optical planes of CNC spindle-shaped rods impart selective reflection of same-handed polarized radiation satisfying the diffraction condition[16,22]

$$\lambda = \bar{n} p \cos \theta$$

where $\lambda$ is the wavelength of reflected radiation, $\bar{n}$ is the average refractive index of the cholesteric medium (taken to be 1.5 for the CNC−organosilica system), $p$ is the pitch of the medium, defined as the distance between two LC building blocks separated by a $2\pi$ rotation, and $\theta$ is the angle of incidence with respect to the helical axis $\chi$ (normal to the plane of the film in our experiments). Cotton CNCs are used in the reflective layers of the sandwich structure due to their smaller size and thus smaller pitch compared to bacterial CNCs, blue-shifting their reflection into the visible and near-IR domains. The diffraction condition dictates reflection centered about a wavelength $\lambda_0 = \bar{n}p$ with a full width at half-maximum $\Delta\lambda = \Delta n p$,[23] with birefringence $\Delta n = (n_e - n_o)$, where $n_e$, $n_o$ are the extraordinary and ordinary refractive indices of the birefringent medium, respectively. Colloidal CNCs preserve the cholesteric assembly upon evaporation in a process known as evaporation-induced self-assembly (EISA),[24,25] forming freestanding lefthanded helicoidal dry films. Post-EISA the films undergo an according pitch shrinkage, blue-shifting the reflection peak of a dried film compared to its aqueous counterpart.[13] Berreman $4 \times 4$ matrix simulation results, detailed below in the numerical modeling section, show reflectivity saturates at a film thickness of ∼$10p$, guiding us to prepare 10−15 μm thick reflective films.

**Experimental Realization and Characterization of Tunable Single-Layer Reflective Films.**
Peak reflection of CNC-based films is readily tuned by increasing organosilica loading from 21.6 wt % to 32.6 wt % to increase its colloidal pitch, red-shifting its reflection wavelength from 400 nm to 915 nm. Freestanding CNC−organosilica films prepared with 21.3 wt % (Table 1) organosilica loading exhibit 45% peak reflection of 400 nm light and appear blue in reflection (Figure 2a), while further increasing their loading to 24.0 wt % (Table 1) resulted in a 41.7% reflection of 557 nm light with a full width at half-maximum of 119 nm, appearing green in reflection (Figure 2b). Further organosilica loading to 26.1 wt % (Table 1) formed a red reflective film (Figure 2c) reflecting 39.5% of 688 nm light with a full width at half-maximum of 122 nm, while loading near the maximum capable of sustaining the cholesteric phase results in a near-IR reflection with a peak reflection of 40.4% at 915 nm (Figure 2d). The films reflect ∼8% of background radiation due to the refractive index contrast of about 0.5 at both optical interfaces of the film. Macroscopic images of a single blue, green, and red reflective film are seen in the inset of Figure 2d, and enlarged images of the green and red reflective films are shown in Figure 2e and f, respectively. The near-IR reflective film features 92% transparency averaged across the visible spectrum (Figure 2g) and excellent flexibility due to the addition of the plasticizer PEG-400 (Figure 2h).

The uniform cholesteric-like CNC layers with magnetic field aligned single-domain organization are characterized by scanning electron microscopy (Figure 3). Maximum reflectivity of a single cholesteric-like CNC film per unit film thickness is attained by alignment of CNCs via subjection to an external static magnetic field during evaporation. CNCs are a negative magnetic susceptibility anisotropic material, and an external magnetic field of magnitude ≥2000 G directed normal to the film's surface can overcome room-temperature thermal fluctuations, aligning the cholesteric LC with its helical axis $\chi$ parallel to the field, with the CNC director orthogonal to the field.[24] Cross-sectional transmission electron micrographs confirm the static magnetic field successfully aligns the cholesteric-like domains of CNC−organosilica composite structures, with domains with pitch as small as 286 nm (Figure 3a) being well aligned, corresponding to blue reflective films. This alignment technique is valid for films reflecting across the visible and near-IR regimes, as macroscopic-sized single-crystalline cholesteric-like domains corresponding to green-reflective films ($p$ = 357 nm, Figure 3b), near-IR reflective films ($p$ = 660 nm, Figure 3c), and films with even larger pitch ($p$ = 1.3 μm, Figure 3d) are realized.

**Cellulose-Based Retarders and Three-Layer Reflective Films.** A tunable, unidirectionally aligned large-scale CNC nematic-like birefringent half-wavelength phase retarder plate is designed to convert incident RCP light into LCP light, nearly doubling reflectivity. It is manufactured by a series of successive linear shear deposition cycles and subsequent solvent evacuation. Shearing colloidal CNCs align the crystallites with the director along the shear direction; this arrangement is then locked into place by rapid evaporation of the colloid's solvent. These films introduce incident radiation to a retardation $L$ related to net birefringence $\Delta n$ and optical path length $d$ given by $L = \Delta n d$. The deposition of successive CNC layers discretely tunes the retardation of the birefringent plate by integer multiples of the retardation of a single layer.

This layer-by-layer manufacturing approach allows for preparation of optical retarders spanning the visible and near-IR ranges of wavelengths. Unidirectionally aligned, nematic-like organization of pure bacterial and cotton CNC films is confirmed by polarized optical and scanning electron microscopies, respectively. Minimal transmitted intensity between crossed polarizers is observed when the shearing direction is parallel to either polarizer (Figure 4a). Transmission between crossed polarizers is maximized when the CNC director is oriented 45° with respect to either polarizer, confirming the presence of unidirectionally aligned CNCs, with their long axes oriented along the shear direction (Figure 4b).

The aligned CNC film features positive birefringence, as evidenced by insertion of a 530 nm retardation plate when the director is oriented +45° with respect to the polarizer, causing the overall system to appear bluish (Figure 4c). Scanning electron microscopy of the shear plane confirms both bacterial CNCs (Figure 4d) and cotton CNCs (Figure 4e) adopt the aligned nematic-like configuration with its director oriented along the shear direction. Nematic-like bacterial CNC-based wave plates with net birefringence $\Delta n = 0.0226$ and retardation spanning 48 to 838 nm (Figure 4f) are formed, and the average retardation per bacterial CNC layer deposited is found to be 54 nm. Bacterial CNCs are larger and retard radiation more than cotton CNCs, enabling course and fine adjustment of the wave plate's retardation. An average wavelength retardation $\Delta nd$ per unit film thickness d for bacterial CNCs is found to be 22.6 nm of retardation per micrometer of film thickness, compared to cotton CNCs at 6 nm of retardation per micrometer of film thickness.

Sandwiching the unidirectionally aligned nematic-like CNC wave plate between two reflective left-handed cholesteric-like CNC films increases reflectivity by as much as 95% compared to a single left-handed reflective film, due to its reflection of all polarizations of incident radiation satisfying the Bragg diffraction condition. Reflection and CD spectra of singlelayer films and their corresponding sandwich structure are recorded following the schematic detailed in Figure 5a. Spectrographical data revealing reflection performance of linearly polarized (LP) radiation show a single cholesteric-like CNC−organosilica composite film reflected only 41.6% of peak radiation (occurring at a wavelength of 558.3 nm), reaching 83.2% of its 50% reflection limit originating from selective reflection due to its left-handed chiral structure (Figure 5b). When LCP light is incident upon the visibly reflective lefthanded single film, a 63% increase of peak reflected radiation compared to incident LP light, as 67.8% reflection of incident 567.0 nm light, is observed (Figure 5b). This case contrasts the case when RCP radiation is incident upon the left-handed single film, in which case no reflection peak is detected (Figure 5b). A ~10% background reflected intensity exists when RCP radiation is incident, due to the refractive index contrast of about 0.5 at both of the film's optical interfaces with air; when this 10% background reflection is subtracted from the LP and LCP curves, we calculate an 83% increase in reflected intensity for incident LCP radiation compared to incident LP radiation. This near doubling of the effective reflection is explained by the nature of LP radiation, which can be thought of as a superposition of in-phase LCP and RCP components. The polarization-dependent reflection of a single left-handed cholesteric-like film contrasts with the reflection of the sandwich structure, where polarization-independent reflection is achieved. A peak reflected intensity of 73.9% of incident LP radiation at 537.4 nm by the sandwich structure is recorded (Figure 5c), corresponding to a 77.64% increase in LP reflection performance compared to a single reflective film. The LP reflective performance of the sandwich structure is similar to its performance when LCP radiation is incident, where a peak reflected intensity of 73.9% of 542.5 nm radiation is captured (Figure 5c). When combined with the RCP reflective performance of the sandwich structure, with a peak reflection of 72.2% at 534.0 nm (Figure 5c), we see the sandwich structure reflects all polarizations of light equally, within experimental error. The 9 nm range of peak reflection of the single-layer film and sandwich structure arises from microstructural defects present within the sample. The reflective CD spectrum of each structure reinforces the contrast seen in the polarization-dependent nature of a single cholesteric-like film compared to the sandwich structure (Figure 5e), as the polarization-dependent single film shows a nontrivial positive CD signal of 60.2% at 567 nm, indicating LCP radiation interacts much more strongly with a single lefthanded film than RCP radiation. The CD signal of the sandwich structure confirms its polarization-independent interaction with light, as a very different CD response is obtained (Figure 5e).

The CD signal of the sandwich structure has a small peak of 11.0% at 567 nm (5.5 times decreased compared to a single film) and is caused by the 9 nm difference in LCP and RCP reflection peaks. Transmission spectra of single films and their corresponding sandwich structure are characterized by using an optical setup schematically shown in Figure 5d.

The corresponding transmission spectra of the single film and sandwich structure show similar transmission of natural light not satisfying the Bragg condition for both structures, except for the reflection trough at 567 nm (Figure 5f), which is found to be 66% lower for the sandwich structure compared to its single film counterpart.

The polarization-dependent interaction of light in a single visibly reflective film and the polarization-independent interaction of light of its corresponding visibly reflective sandwich structure are confirmed and visualized via reflective optical microscopy. A single reflective film (inset of Figure 6a) is seen to moderately reflect green LCP light (Figure 6a), while only marginally reflecting incident green LP light (Figure 6b). When RCP radiation is incident upon a single reflective film, about 10% of all light across the visible spectrum is reflected, and so the corresponding micrograph is dark gray (Figure 6c). The sandwich structure (inset of Figure 6d) reflects light with all polarizations of incident radiation equally, as reflective micrographs captured after LCP radiation is incident (Figure 6d) have similar color and intensity to micrographs obtained after LP (Figure 6e) and RCP (Figure 6f) radiation is incident. These micrographs of reflective color are consistent with the reflective spectra presented earlier (Figure 5).

Visibly transparent, near-IR reflective CNC−organosilica photonic structures are prepared by increasing organosilica loading to red-shift both helicoidal optical components' reflection. The polarization-dependent nature of a single film compared to the polarization-independent reflection of its corresponding sandwich structure emulates our observations of the visibly reflective photonic crystal structures. Their reflection, CD, and transmission spectra are obtained using the optical setup detailed in Figure 5a and d, respectively. A single reflective film has peak LP reflectivity 40.4% at 915.4 nm (Figure 7a) and comparatively shows a 97.0% increase in reflectivity when LCP radiation is incident, reaching 79.6% reflection at 915.4 nm (Figure 7a). When RCP radiation is incident upon a single reflective film, no reflection peak is observed and an average background reflectivity of 6.4% is captured (Figure 7a) due to reflection caused by the refractive index contrast at each of the film's optical interfaces with air. The IR-reflective sandwich structure's reflection is polarization independent, as a peak reflectivity of 83.6% at 895.6 nm is seen when LCP radiation is incident, comparable to peak reflectivity of 81.9% at the same wavelength when LP radiation is incident and peak reflectivity of 80.3% at 893.3 nm when RCP radiation is incident (Figure 7b). The CD signal of the sandwich structure confirms its polarization-independent interaction with light, as a weak CD signal is obtained, averaging 2.1% (Figure 7c). A single reflective film's CD spectrum shows a positive signal peaking at 72.5% at 915.4 nm, revealing its highly polarization-dependent interaction with incident radiation (Figure 7c). Visible transmission of the IR-reflective single film and sandwich structure yields comparable results, with a single film transmitting an average of 90.1% of incident visible radiation and the sandwich structure transmitting an average of 85.0% of visible radiation (Figure 7d). The 5.1% difference between these two curves arises from the doubled number of optical interfaces of the sandwich structure compared to a single reflective film. Visibly transparent IR reflectors are of particular interest in applications such as solar gain regulators, radiative heat filters, microscopy optical elements, and other applications where optical observations need to take place alongside thermal shielding.

**Numerical Modeling.** Known analytical approximations[23] do not take into account many of the experimental features present in our cholesteric assemblies and absorption of the material. To properly model propagation of light through our cholesteric-like structures, we use numerical modeling based on the Berreman 4 × 4 matrix method.[26,27] This method allows one to calculate the reflectance of our cholesteric-like film in a broad spectral range while also taking into account cellulose's minimal absorption. The results of these numerical simulations are presented in Figure 8. Our numerical models are consistent with our experimental finding that the three-layer photonic structure allows for the near-100% reflection of incident light due to their reflection of both handedness of circularly polarized radiation (Figure 8a).

Reflective efficiency depends on the thickness of the cholesteric-like films, which we cast in terms of the d/p ratio, and also can be quantified by means of the integral reflectance $R_\Sigma = \int_{\lambda_1}^{\lambda_2} R(\lambda)d\lambda$, where $R(\lambda)$ is the reflectance spectrum within a range $\lambda_1$-$\lambda_2$ at a given *d/p*. Numerical calculations show that reflective efficiency increases exponentially with the thickness of the film until it saturates at a film thickness equal to about 10 cholesteric pitches (Figure 8b), guiding our experimental efforts. The number of cholesteric pitches needed to saturate reflectance also depends on $\Delta n$ and increases as $\Delta n$ decreases. Calculations show that the reflective efficiency from the sandwich structure depends on the retardation of the CNC retardation plate and is maximized when $\Delta nd \approx \lambda/2$ (Figure 8c). Our modeling reveals that the polarization-independent selective reflectivity of light can be designed to occur within a narrow part of the visible or near-IR spectral ranges (Figure 8a) or over a broader spectral range when cholesteric layers have nonzero pitch gradients across the film thickness. Importantly, this modeling also agrees with our experiments showing IR reflective films can be visibly transparent and, conversely, visibly reflective films can be IR-transparent, which is essential for many applications, including retrofittable window films for solar gain regulation.

**DISCUSSION**

Our findings illustrate that cholesteric-like photonic structures with high reflectivity can be obtained using the abundant, inexpensive biomaterial cellulose. The CNC-based tunable reflective sandwich films strongly interact with all polarizations of light, reflecting up to 83.6% of incident unpolarized radiation at wavelengths fulfilling the Bragg diffraction condition (with a full width at half-maximum of about 120 nm), and are more than 85% transparent to radiation outside this range. Noteworthy, each CNC-based optical element composing the sandwich structure also serves as an individual optical element and may find separate applications. For example, nematic-like bacterial and cotton CNC assemblies operate as solid birefringent retarder plates with predesigned tunable phase retardation, which could be of interest for applications in polarizing optics and birefringent compensators used in liquid crystal displays. Preparing this structure entailed forming tunable, uniaxially aligned nematic-like CNC retardation plates via a series of repeated linear shear deposition cycles, kinetically preventing the CNC rods from self-assembling into their native cholesteric phase. Meanwhile, cholesteric-like dried CNC films function as tunable LCP light reflectors and as right-handed circular polarizers.

Self-assembly of nanocellulose into ordered optical structures increases their applied scope compared to conventional ordered nanostructures produced using top-down methods and, in doing so, promotes implementation in applications as diverse as smart clothes, document authenticity features, tunable inks, and visible-range transparent low-emissivity films suitable for retrofitting windows to boost their energy efficiency. From the standpoint of the latter application, various climates have different needs in terms of infrared radiation transmission and reflection to maximize building energy efficiency. In hot climates, it is desirable to reflect the sun's predominately near-IR radiation, thereby saving on air conditioning costs. Our near-IR reflecting films are suitable for this purpose. On the other hand, buildings in colder climates may save energy by letting the sun's near-IR radiation transmit through while benefiting from reflecting and trapping in the room's emission due to room-temperature blackbody radiation. Reflection of this predominantly mid-IR radiation (peaked at about 10 μm) could save energy and reduce the cost needed for building heating. Our current approach will need to be developed further to serve this need. During EISA, the colloidal CNCs dispersion's pitch shrinks about 90%, from ~6 μm in the colloidal dispersion (with the exact pitch depending on CNC concentration, surface charge density, and processing conditions) to about 600 nm once fully dried. Therefore, a different approach, which could lock the cholesteric pitch at 6 μm or so, is needed to fabricate mid-IR and other longer-wavelength reflective films. Such cholesteric-like mid-IR reflective films can be developed too, which will be reported elsewhere. In general, our cellulose-based multilayer films are suitable candidates for applications such as low-emissivity solar-gain-regulating window films that can be predesigned to give spectral reflectivity optimized for either hot or cold climates without compromising visible-range optical transparency.

**CONCLUSIONS AND OUTLOOK**

In this work we have numerically modeled, experimentally implemented, and optically characterized transparent and flexible laminar composite cholesteric- and nematic-like CNC based photonic structures with reflection tunable from visible to near-IR. We have shown that this sandwich structure features tunable, polarization-independent reflection approaching 100% reflectivity, nearly double that compared to its corresponding single cholesteric-like CNC film within the visible and near-IR spectral ranges. The nematic-like CNC films' birefringence can be coarsely and finely tuned with bacterial and cotton CNC deposition. Such low-cost, highly reflective films made from renewable cellulose sources may find applications in thermally isolating viewports, economical color filters, 3D-printable pigments, document security features, and solar-gain-regulating technologies. Furthermore, its reflection may be further redshifted into the mid-IR regime by locking in its pitch prior to EISA, useful to trap heat in colder climates while still serving as visibly transparent window-retrofitting films with controlled solar gain.

**METHODS**

Bacterial cellulose pellicles served as one of our renewable cellulose sources to form CNCs and are obtained using *Acetobacter hansenii* cultured in Hestrin-Schramm medium at 26 °C for 3 weeks with a dry mass yield of 7.5 g/L. In situ darkfield microscopy reveals cellulose is formed at solid/liquid and gas/liquid interfaces and that the bacteria excrete cellulose along their length. During this process neighboring secreted nanofibers hydrogen bond to one another, forming ribbon-shaped cellulosic assemblies hundreds of micrometers in length, with a cross-sectional area of ∼10 nm by ∼50 nm.[28] Cellulose production also serves as a bacteria propulsion mechanism, and cellulose-ejecting bacteria move as fast as 6 μm/min in the direction opposite the newly ejected cellulosic ribbon. Other bacteria can glide on established cellulose ribbons (Figure 9a) at about the same speed. After cellulose secretion, bacteria detach from the cellulose fibrils (Figure 9b). At some critical cellulosic ribbon concentration, the ribbons begin to physically cross-link via hydrogen bonds, forming an extended network (Figure 9c) and eventually an entire cellulose pellicle.

Colloidal CNCs are prepared from bacterial cellulose and cotton balls (Swisspers brand). Raw bacterial cellulose pellicles are purified by soaking in 1 wt % sodium hydroxide solution at 70 °C for 1 h, followed by extensive washing with distilled water and drying at ambient conditions. The cotton balls are used without further purification. The cellulose source is hydrolyzed to yield CNCs by modifying the protocol in the literature.[13] Briefly, the cellulose is mechanically stirred at 1000 rpm in 64 wt % sulfuric acid at 45 °C with an optimized cellulose-to-acid mass ratio of 1:25 for a preset time (1 h for bacterial cellulose and 2 h for cotton cellulose, respectively) assisted by continuous sonication in a bath sonicator (Branson 3800). After the reaction, CNCs are decorated with negatively charged sulfate ester moieties substituting about 5% of primary alcohols present along the CNCs' main chain backbone (Figure 9d).[29] Colloidal CNCs are then purified by aqueous dilution and repeated washing via centrifugation at 10k g-force for 8 min intervals. The supernatant is then dialyzed against deionized water using a membrane with a cutoff molecular weight of 12 000−14 000 g/mol until the pH of the dialysis water reached 7, followed by brief sonication, centrifuged at 10 000 g-force for 8 min one last time to remove remaining microsized CNC aggregations, and then finally filtrated through 11 μm pores (Whatman no. 1 filter paper) for bacterial CNCs and 5 μm pores (Millipore MF filters) for cotton CNCs. The study detailed herein utilizes both bacterial- and cotton-derived nanocellulose, with CNCs measuring approximately 50 nm by 2 μm (Figure 9e) and 10 nm by 200 nm (Figure 9f), respectively.

Composite cotton CNC−organosilica solid reflective films are prepared by modifying the protocol previously reported in the literature,30 using organosilica precursor 1,2-bis-(trimethoxysilyl)ethane (96% pure, Alfa Aesar), and with addition of polyethylene glycol (number-average molecular weight 400) prior to 1 h of magnetic stirring at 600 rpm of the organosilica precursor with aqueous CNCs. The film is cast on a polystyrene Petri dish and cured in ambient conditions in the presence of a 2000 G magnetic field oriented normal to the film surface, to align the helical axis uniformly perpendicular to the film surface. After evaporation, the solid freestanding film preserves the cholesteric-like configuration with an accompanying pitch shrinkage. The four reflective cholesteric-like CNC− organosilica films detailed herein have the composition listed in Table 1.

The solid nematic-like bacterial or cotton CNC retardation plate is prepared by manual linear shear deposition of aqueous 3.5−5.0 wt % bacterial CNCs or 3.5−5.0 wt % cotton CNCs on a glass plate or directly on top of a cholesteric-like CNC film at a rate of 0.25 cm/s, followed by drying on a 30−35 °C hot plate.[31] This process is repeated until the desired half-wavelength retardation is achieved. When shear depositing on a glass plate, the nematic-like retarder film was removed from the glass plate by shearing with a cutting edge, leaving a freestanding film. All sandwich structures detailed herein were prepared by linear shear deposition of aqueous CNC directly on one cholesteric-like CNC-based film. To form the final sandwich structure from this two-layered film, another single cholesteric-like CNC-based reflective film is adhered to the two-layered film, with the nematic-like layer in the center, using a thin layer of adhesive (NOA 65, Norland Products, Inc.), followed by 1 min of 365 nm wavelength curing (OmniCure Series 2000).

CNCs' morphology and geometry are elucidated by transmission electron microscopy (FEI Philips CM100) operating at 80 kV. The uniaxially aligned nematic-like CNC layer's structure is determined via scanning electron microscopy (Carl Zeiss EVO MA 10) operating at 5 kV. Optical transmission, reflection, and CD spectra are obtained at normal incidence with an Ocean Optics USB2000+ spectrometer in conjunction with an Olympus BX-51 microscope equipped with a linear polarizer, an appropriate achromatic quarter-wave plate (Thorlabs, Inc., AQWP05M-600 or AQWP05M-980), and an Olympus LMPLFLN 20× objective. Polarized optical micrographs are captured via CCD (Point Grey Research, GS3- U3-28S5C). CNC orientation in the optical retardation layer is deduced with the insertion of a 530 nm retardation plate, and its retardation is quantified with a Berek compensator (Olympus U-CTB).

ACKNOWLEDGMENTS

This research was supported by the U.S. Department of Energy, Advanced Research Projects Agency-Energy award DEAR0000743. We thank Eugenia Kumacheva for insightful discussions and David A. Rudman and Yao Zhai for technical assistance.

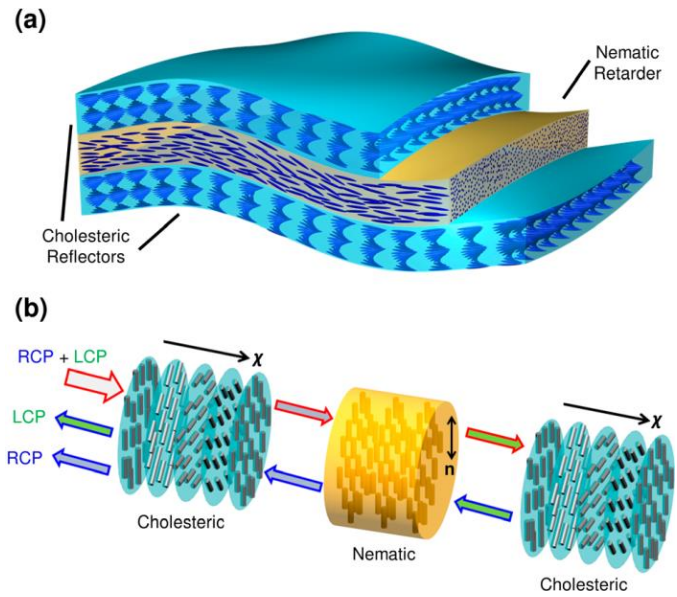

**Figure 1.** Schematics detailing (a) the proposed sandwich structure's architecture and flexibility and (b) its cross-section depicting the mechanism enabling its polarization-independent selective Bragg reflection. The small rods represent individual CNCs (not to scale).

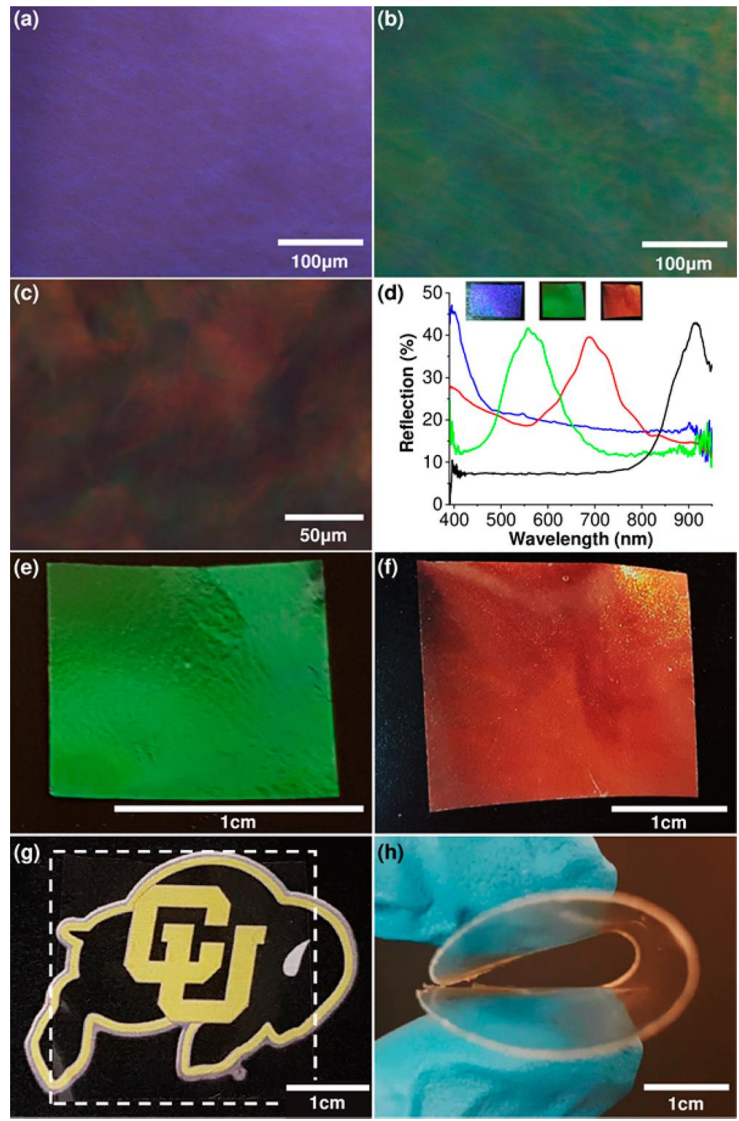

**Figure 2.** Reflective optical micrographs with natural light incident upon CNC-based reflective films with reflection peaks centered at (a) 400 nm (b) 557 nm, and (c) 688 nm prepared with organosilica loading ranging from 21.3 to 32.6 wt % in the presence of a static 2000 G magnetic field directed parallel to the surface normal. In (d) these films' reflection spectrum is plotted in their predominant reflective color. The near-IR reflective film's reflectivity peaks at 915 nm and is plotted in black. In the inset, a 2 cm wide blue, 1.15 cm wide green, and 2 cm wide red reflective cholesteric-like film are shown. (e) Image of the green-reflective cholesteric-like film. (f) Image of the red-reflective cholesteric-like film. In (g) a highly visibly transparent near- IR reflective film is shown, with its perimeter outlined using white dashed marks. (h) Demonstration of flexibility of the film enabled by the addition of 30 wt % of the plasticizer PEG-400.

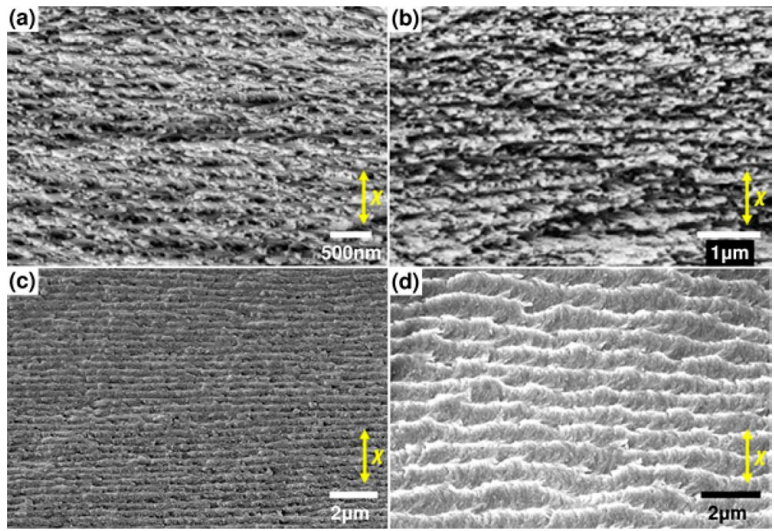

**Figure 3.** Cross-sectional scanning electron micrographs of dried CNC-based cholesteric-like films confirm effectiveness of static 2000 G magnetic field alignment in promoting macroscopic single-domain configurations. The pitch, and thus the corresponding reflection peak, is tunable from $p =$ (a) 286 nm to (b) 357 nm to (c) 660 nm to (d) 1.3 µm.

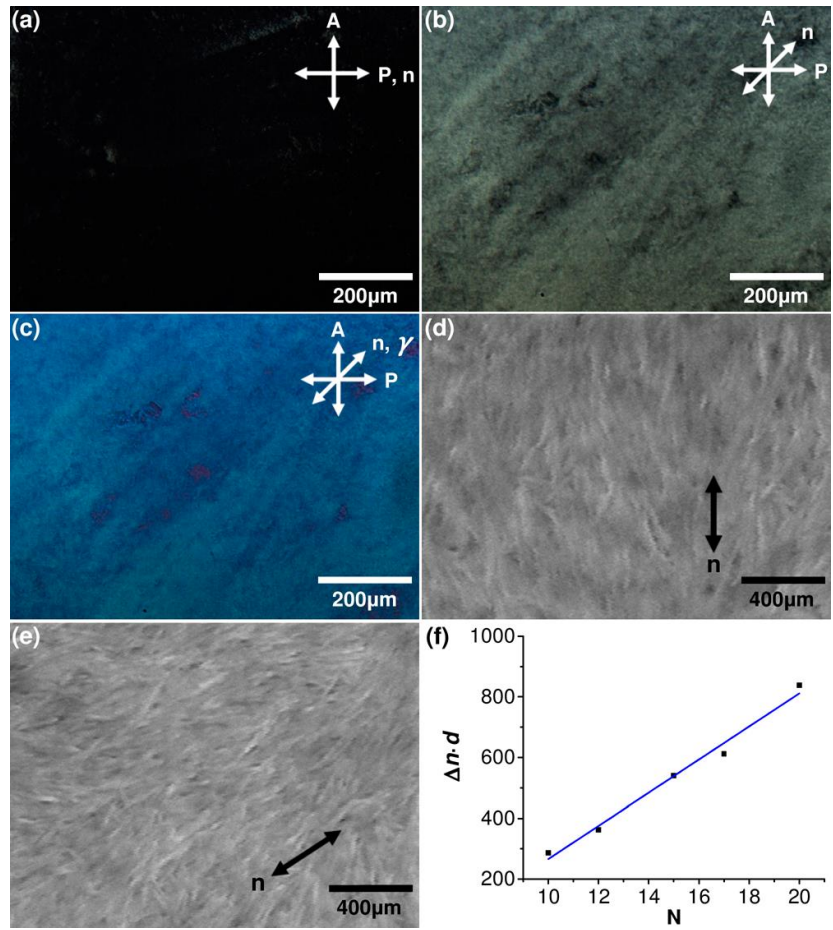

**Figure 4.** Polarized optical micrographs of the CNC-based retardation plates with the director oriented (a) parallel to the polarizer and (b, c) +45° with respect to the polarizer (b) before and (c) after insertion of a 530 nm retardation plate. Scanning electron micrographs taken along the direction normal to the shear plane for the films with both (d) bacterial CNCs and (e) cotton CNCs, showing unidirectional alignment along the shear direction. The CNC director has orientation denoted by the double-headed arrow labeled **n**. (f) Linear relation of $\Delta nd$ vs the number of deposited CNC layers $N$ of a unidirectionally aligned nematic-like bacterial CNC wave plate.

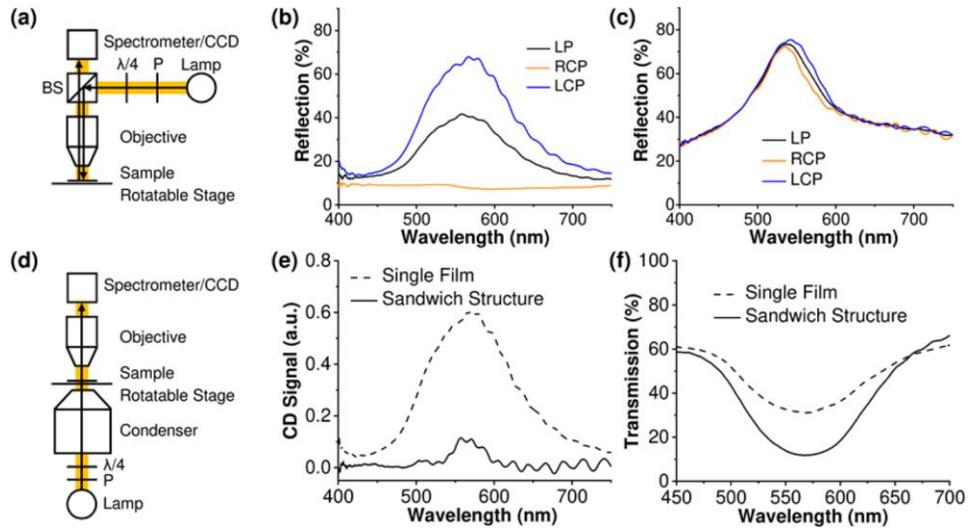

**Figure 5.** (a) Optical setup used to characterize reflection and CD spectra of both single-layer reflective films and their corresponding sandwich structure. P is a linear polarizer, *λ/4* is an achromatic quarter-wave plate, and BS is a beam splitter. (b) Optical characterization of a visible-range single reflective film and its corresponding sandwich structure. The reflection spectrum of a single reflective cotton-based CNC-organosilica film with LP, RCP, and LCP incident light. (c) Corresponding reflection spectra of a visibly reflective sandwich structure. (d) Optical setup used to characterize transmission spectra of both single reflective films and their corresponding sandwich structure. (e) CD spectra of both the single film and the sandwich structure. (f) Transmission spectra with natural light incident on a single film and its analogous sandwich structure.

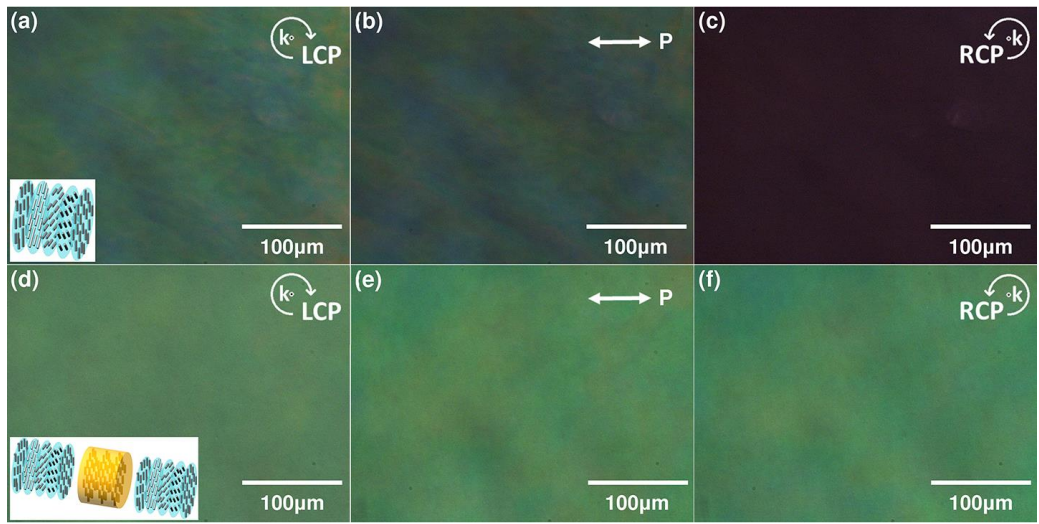

**Figure 6.** Reflective optical micrographs of a single composite cotton CNC-organosilica

reflective film (top row) with incident (a) LCP, (b) LP, and (c) RCP radiation and of its corresponding sandwich structure (bottom row), as seen with the incident (d) LCP, (e) LP, and (f) RCP light. LCP and RCP are respectively denoted by clockwise and counterclockwise circular arrows centered about a wave vector k parallel to $\chi$.

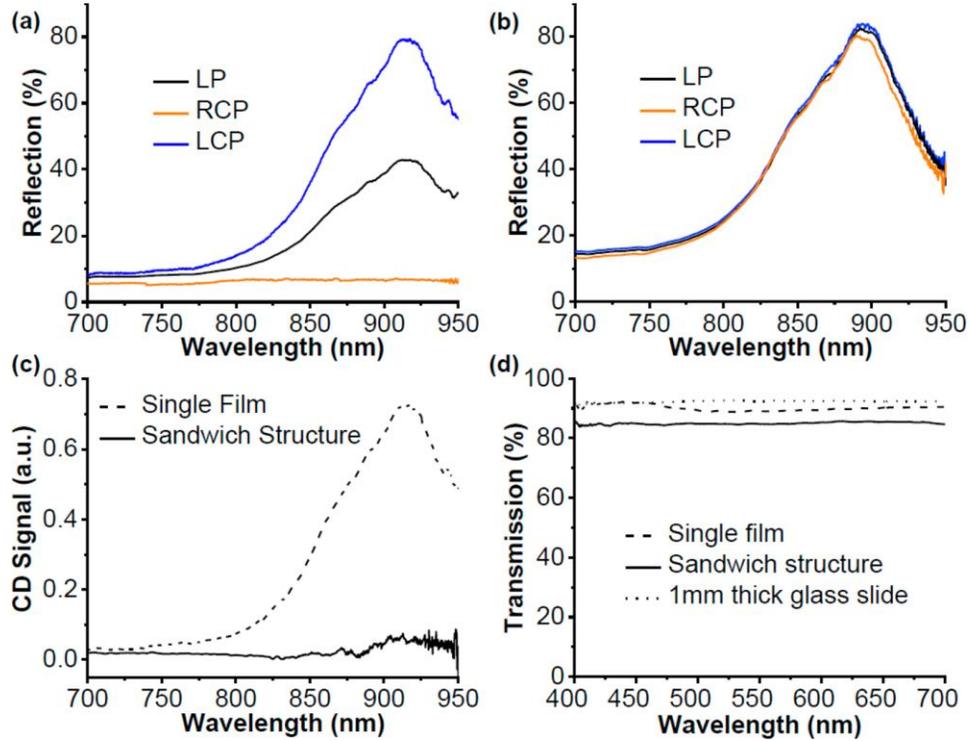

**Figure 7.** Optical characterization of a visibly transparent, single near-IR reflective film and its corresponding sandwich structure. (a) Reflection spectrum of a single reflective cotton CNC-organosilica film with incident LP, RCP, and LCP radiation. (b) Corresponding reflection spectra of a sandwich structure. (c) CD spectra of both the single film and the sandwich structure. (d) Transmission spectra with natural light incident upon a single film and its corresponding sandwich structure, showing average visible transmission of 90.1% for a single reflective film and 85.0% for the corresponding sandwich structure. The transmission of a 1 mm thick glass slide, averaging 92.3% visible transmission, is provided for comparison.

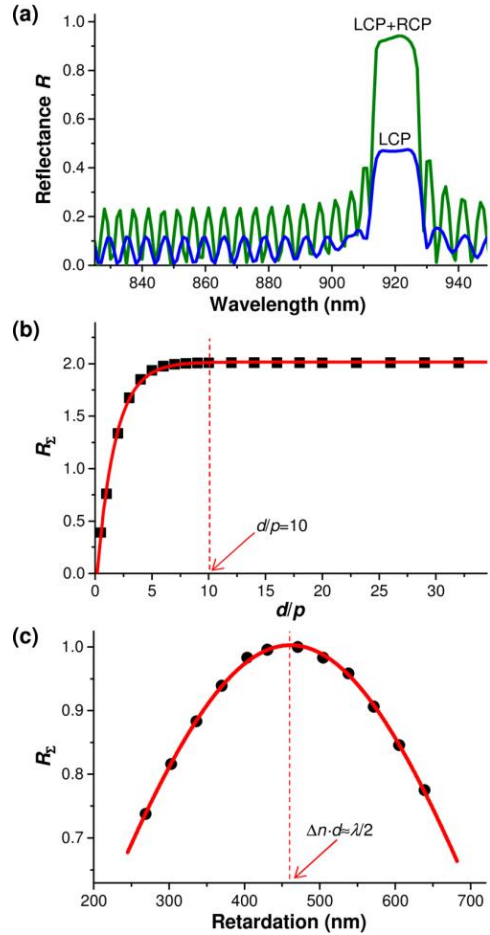

**Figure 8.** (a) Computer-simulated reflection for near-IR reflective photonic structures, showing a theoretically doubled reflection of a sandwich structure (plotted in green) as compared to its single film counterpart (plotted in blue). (b) Reflective efficiency's dependency on the film thickness-to-pitch (d/p) ratio. (c) Reflective efficiency of the three-layer composite structure shown in (a) depends on the effective retardation of the CNC retardation plate and is maximized when $\Delta nd \approx \lambda/2$.

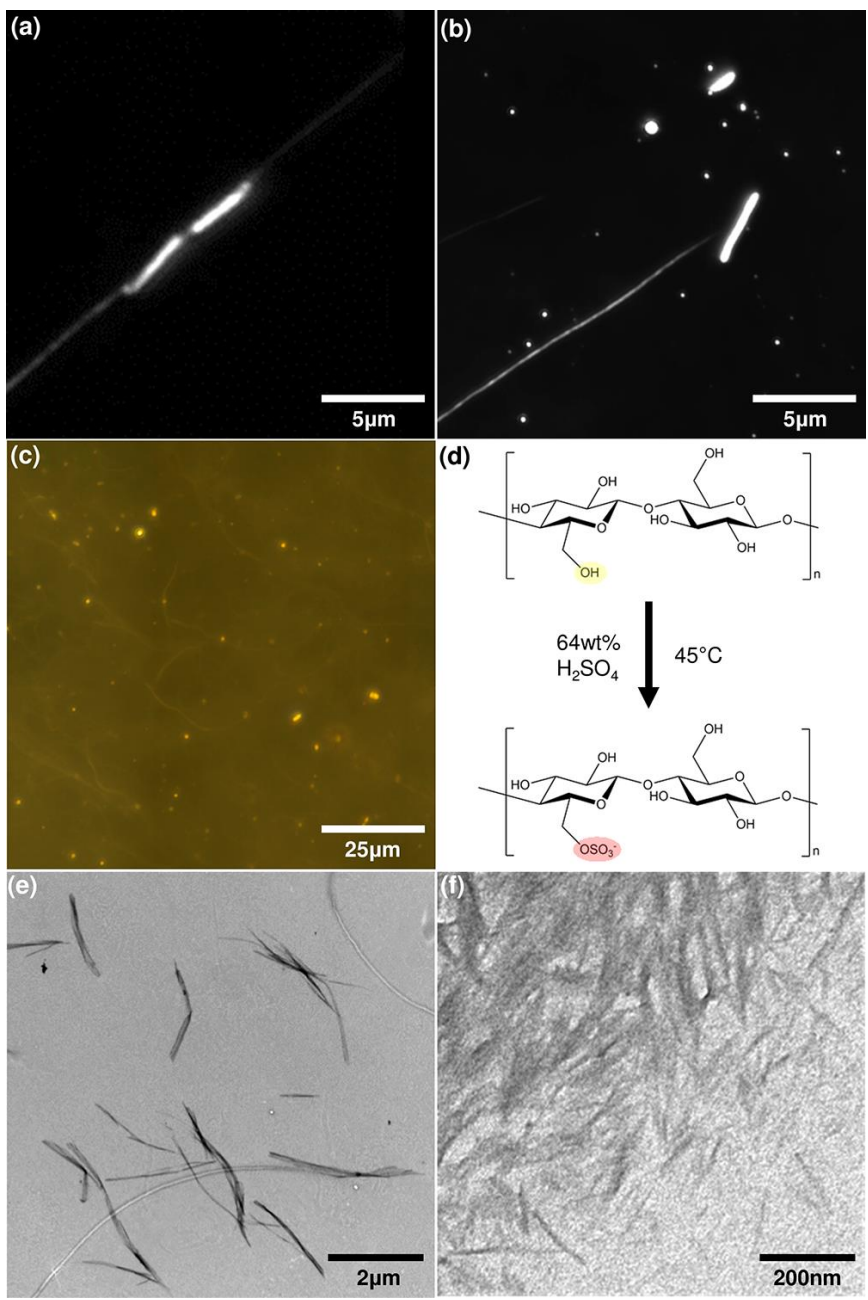

**Figure 9.** Dark-field micrographs of bacterium *Acetobacter hansenii* (a) traversing along a cellulose fibril and (b) detaching from an excreted cellulose ribbon, freely moving within the culture medium. (c) Several *Acetobacter hansenii* bacteria embedded within the loosely physically cross-linked cellulosic network. (d) During CNC synthesis, some portion of the cellulose's primary alcohols (located on C6, highlighted in yellow) are sulfated, leaving the rod with a net negative charge (highlighted in red). Transmission electron micrographs of (e) bacterial CNCs and (f) cotton CNCs.

Table 1. Composition of CNC-Organosilica Reflective Films with Predesigned Reflective Colors

| Component | Reflective color | | | |
| --- | --- | --- | --- | --- |
| | Blue | Green | Red | Near-IR |
| cotton CNC (wt %) | 48.7 | 46.0 | 43.9 | 37.4 |
| organosilica (wt %) | 21.3 | 24.0 | 26.1 | 32.6 |
| PEG-400 (wt %) | 30.0 | 30.0 | 30.0 | 30.0 |